\documentclass[letterpaper,aps,pra,twocolumn,showpacs,superscriptaddress,amsmath,amssymb]{revtex4-1}

\usepackage{graphicx}
\usepackage[latin1]{inputenc}
\usepackage{dcolumn}
\usepackage{hyperref}
\usepackage{color}
\usepackage{mathrsfs}
\usepackage{bm}

\begin{document}

\newcommand*{\ii}[0]{\mathrm{i}}
\newcommand*{\dd}[0]{\mathrm{d}}
\newcommand*{\e}[0]{\mathrm{e}}
\newcommand*{\tq}[1]{\left.#1\right|_{0}}
\newcommand*{\tr}[0]{\mathrm{tr}}
\def\vec#1{\mathbf{#1}}
\def\ket#1{|#1\rangle}
\def\bra#1{\langle#1|}
\def\ketbra#1{|#1\rangle\langle#1|}
\def\braket#1{\langle#1|#1\rangle}
\def\idmat{\mathbf{1}}
\def\caln{\mathcal{N}}
\def\calc{\mathcal{C}}
\def\rhon{\rho_{\mathcal{N}}}
\def\rhoc{\rho_{\mathcal{C}}}
\def\tr{\mathrm{tr}}
\def\bfu{\mathbf{u}}
\def\bfmu{\mbox{\boldmath$\mu$}}
\def\bfxi{\mbox{\boldmath$\xi$}}

\newcommand{\ri}{{\rm i}}
\newcommand{\re}{{\rm e}}
\newcommand{\bb}{{\bf b}}
\newcommand{\bc}{{\bf c}}
\newcommand{\bx}{{\bf x}}
\newcommand{\bz}{{\bf z}}
\newcommand{\by}{{\bf y}}
\newcommand{\bv}{{\bf v}}
\newcommand{\bd}{{\bf d}}
\newcommand{\br}{{\bf r}}
\newcommand{\bp}{{\bf p}}
\newcommand{\bq}{{\bf q}}
\newcommand{\bk}{{\bf k}}
\newcommand{\bA}{{\bf A}}
\newcommand{\bB}{{\bf B}}
\newcommand{\bE}{{\bf E}}
\newcommand{\bF}{{\bf F}}
\newcommand{\bH}{{\bf H}}
\newcommand{\bI}{{\bf I}}
\newcommand{\bR}{{\bf R}}
\newcommand{\bM}{{\bf M}}
\newcommand{\bX}{{\bf X}}
\newcommand{\bXo}{{\overline{\bf X}}}
\newcommand{\bxo}{{\overline{\bf x}}}
\newcommand{\bn}{{\bf n}}
\newcommand{\bs}{{\bf s}}
\newcommand{\tbs}{\tilde{\bf s}}
\newcommand{\rSi}{{\rm Si}}
\newcommand{\dB}{d_{\rm Bures}}
\newcommand{\beps}{\mbox{\boldmath{$\epsilon$}}}
\newcommand{\bthe}{\bm{\theta}}
\newcommand{\blam}{\bm{\lambda}}
\newcommand{\rg}{{\rm g}}
\newcommand{\xmax}{x_{\rm max}}
\newcommand{\ra}{{\rm a}}
\newcommand{\rx}{{\rm x}}
\newcommand{\rs}{{\rm s}}
\newcommand{\rP}{{\rm P}}
\newcommand{\up}{\uparrow}
\newcommand{\down}{\downarrow}
\newcommand{\hc}{H_{\rm cond}}
\newcommand{\kb}{k_{\rm B}}
\newcommand{\cI}{{\cal I}}
\newcommand{\tit}{\tilde{t}}
\newcommand{\cE}{{\cal E}}
\newcommand{\cC}{{\cal C}}
\newcommand{\Ubs}{U_{\rm BS}}
\newcommand{\qq}{{\bf ???}}
\newcommand*{\etal}{\textit{et al.}}

\newcommand{\cor}{{\bf !!!}}
\newcommand{\corr}{{\bf ?`?`?`}}

\title{Quantum parameter estimation using general single-mode Gaussian states}

\author{O. Pinel}
 \affiliation{Centre for Quantum Computation and Communication Technology, Department of Quantum Science, \\
The Australian National University, Canberra, ACT 0200, Australia}
\author{P. Jian}
 \affiliation{Laboratoire Kastler Brossel,  Universit\'e Pierre et Marie Curie-Paris 6,\\
ENS, CNRS; 4 place Jussieu, 75252 Paris, France}
\author{N. Treps}
 \affiliation{Laboratoire Kastler Brossel,  Universit\'e Pierre et Marie Curie-Paris 6,\\
ENS, CNRS; 4 place Jussieu, 75252 Paris, France}
\author{C. Fabre}
 \affiliation{Laboratoire Kastler Brossel,  Universit\'e Pierre et Marie Curie-Paris 6,\\
ENS, CNRS; 4 place Jussieu, 75252 Paris, France}
\author{D. Braun}
 \affiliation{Laboratoire de Physique Th\'eorique, Universit\'e Paul Sabatier, Toulouse III and CNRS, \\
118 route de Narbonne, 31062 Toulouse, France}

\date{\today}

\begin{abstract}

We calculate the quantum Cram\'er--Rao bound for the sensitivity with which
one or several parameters, encoded in a general single-mode Gaussian state,
can be estimated. 
This includes in particular the interesting case of mixed Gaussian
states. We apply the formula to the problems of estimating phase, purity,
loss, amplitude, and squeezing. In the case of the simultaneous measurement
of several 
parameters, we provide the full quantum Fisher information matrix.  Our
results unify previously known partial results, and constitute a complete
solution to the problem of knowing the best possible sensitivity of
measurements based on a single-mode Gaussian state.
\end{abstract}

\maketitle

Metrology using electromagnetic fields as a probe is of fundamental
importance in 
many areas of science and technology. Applications include, amongst many
others, 
distance measurements with laser range finders or radar, measurement of the
shape and composition of objects in microscopy and spectroscopy, angular
velocities 
with laser gyroscopes, and attempts of gravitational wave detection using large
interferometers such as VIRGO and LIGO. In all these schemes, one or several parameters of
the system under investigation are encoded in the state of light, and one
subsequently tries to recover that value by detecting the light in a suitable
way. It is important to know with what precision such a parameter can be
measured in principle, i.e.~once all technical noise sources are eliminated,
measurement instruments are ideally precise, and the system can be prepared in
the same identical state as often as desired \cite{Wiseman2009}.

Quantum parameter estimation theory provides an answer to this question in the
form of the quantum Cram\'er--Rao bound, which constitutes a lower bound to the
fluctuations of an estimator of a parameter $\theta$, given the knowledge of how
the quantum mechanical state $\rho$ depends on the parameter. The bound is
essentially due to quantum uncertainty and given by the inverse quantum Fisher
information $I_{\rm Fisher}$ associated with the state $\rho_{\theta}$, where
$I_{\rm Fisher}$ measures the distinguishability (or, in a complementary way,
the fidelity) of two close-by quantum states that differ infinitesimally in
$\theta$. The result can be intuitively understood in quantum information
terms. For neighbouring states that differ slightly in the value of a parameter $\theta$, the more distinguishable the states, the more precisely $\theta$ can be measured. The quantum Cram\'er--Rao bound is applicable
to any quantum mechanical system and provides often a generalized uncertainty
relation, even if no Hermitian operator can be simply associated with a given
observable, as is the case for example for phase estimation
\cite{Helstrom1969,Braunstein1994,Braunstein1996}.

In quantum optics, a particularly useful class of states are Gaussian states,
which are defined generally as states with a Gaussian Wigner function. This
class includes coherent states (e.g.~the light emitted by a laser operating far
above threshold), thermal light, squeezed light, and, in the case of several
modes, some entangled states such as EPR states. These states are readily available in the laboratory
with large photon numbers \cite{Keller2008} and play an important role in quantum
metrology and information processing \cite{Wang2007}. In \cite{Pinel2012}
quantum Fisher information was calculated for pure Gaussian states with
arbitrarily many modes, and a measurement scheme was proposed that saturates
the quantum Cram\'er--Rao bound. However, the need to calculate the square root
of two different operators renders the calculation in general very difficult for
mixed states of infinite dimensional systems. Partial early results include
those by Twamley et al.~who calculated the Bures distance between squeezed
thermal states \cite{Twamley1996}, and Paraoanu et al.~who did so for displaced
thermal states \cite{Paraoanu1998}. Scutaru found the fidelity for thermal
states that are both displaced and squeezed \cite{Scutaru1998}. Monras and
Paris 
found the quantum Fisher information for the particular problem of loss
estimation with displaced squeezed thermal states \cite{Monras2007}, and
Aspachs 
et al.~considered phase estimation with thermal states
\cite{Aspachs2009}. These 
results all refer to single-mode states. Very recently Marian and Marian
produced a result for the fidelity between arbitrary one- or two-mode Gaussian
states \cite{Marian2012}.

Here we provide a comprehensive analysis for general single-mode Gaussian states.
They can be parameterized by five real parameters that we will describe below. Our
analysis is based on a general expression for the Bures distance between two
Gaussian one-mode states in \cite{Scutaru1998}, which we expand up to second
order in the infinitesimal difference $d\theta$ in the parameters between the
two neighboring states $\rho_\theta$ and $\rho_{\theta+d\theta}$. This yields
the quantum Fisher information. For the case of simultaneous estimation of
several parameters, we calculate the complete quantum Fisher matrix, which sets
a lower bound to the covariance matrix of the parameters in the sense of a
matrix inequality \cite{Paris2009}.

{\em Gaussian states.}
The quadratures of an electromagnetic field mode (in units with $\hbar=2$) 
 are
defined in terms of the annihilation and creation operators $a$ and $a^\dagger$
of the mode as \cite{Scully1997}
\begin{equation} \label{xipi}
\hat{x}=a^\dagger+a\mbox{,
}\hat{p}=\ii(a^\dagger-a)\,.
\end{equation}
In the Wigner function description of the state, the quadratures correspond
to two phase space coordinates $x$ and $p$ which we group into a 2D vector
$\bX$, 
$\bX^\top=(x,p)$. The Wigner function for an arbitrary
quantum state given in  terms of its density matrix $\rho$ is then defined as
\begin{equation} \label{Wdef}
W(x,p)=\frac{1}{2\pi}\int_{-\infty}^\infty d\xi e^{-i p
  \xi}\langle x-\xi|\rho|x+\xi\rangle\,. 
\end{equation}
For a single-mode Gaussian state that depends on the parameter $\theta$, the
Wigner function takes the general form
\begin{equation} \label{Wg}
W_\theta(\bX)=\frac{1}{2\pi|\det
\Sigma_\theta|^{1/2}}e^{-\frac{1}{2}(\bX-\bXo_\theta)^\top\Sigma_\theta^{-1}(\bX-\bXo_\theta)}\,,
\end{equation}
where $\bXo_\theta$ are the parameter dependent expectation
values of the quadratures in the state $\rho_{\theta}$, and $\Sigma_\theta$ is
the covariance matrix \cite{Weedbrook2012}. The latter is a real symmetric
matrix with matrix 
elements
\begin{equation} \label{gij}
\Sigma_{\theta,ij}=\frac{1}{2}\langle X_i X_j+X_j X_i\rangle- \langle X_i \rangle \langle X_j\rangle \,,
\end{equation}
and $\langle\ldots\rangle\equiv\tr(\rho\ldots)$. We see that the Wigner function
is parameterized with five real parameters.
The purity of the state is given by  $P_\theta=\tr \rho_\theta^2=(\det\Sigma_\theta)^{-1/2}$.

{\em Quantum Cram\'er--Rao bound.}
The (squared) sensitivity $(\delta\theta)^2$ with which a parameter $\theta$ can
be estimated from $Q$ measurement results $a_i$ of some observable $A$ is
defined as the variance of the deviation from the true value of $\theta$ of an estimator of $\theta$, $\theta_{\rm est}(a_1,\ldots,a_Q)$, that
depends solely on the measurements results: $\delta\theta^2=\langle(\theta_{\rm{est}}(a_1,\ldots,a_Q)-\theta)^2\rangle_s$ where $\langle\ldots\rangle_s$ corresponds to the statistical mean. It is bounded from below by the inverse of the quantum 
Fisher information,
\begin{equation}
  \label{eq:QCR}
  (\delta\theta)^2\ge\frac{1}{QI_{\rm Fisher}(\rho_\theta)}\,.
\end{equation}
where $I_{\rm Fisher}$ is defined here as the quantum Fisher information for a single measurement. The bound is optimized over all possible POVM measurements and classical
post-processing of data (i.e.~all estimator functions). For an unbiased estimator it can
be saturated in the limit of a large number of measurements, and thus
represents 
the ultimate reachable bound of sensitivity. The quantum Fisher information is
given in terms of the Bures distance between two close-by states
$\rho_{\theta},\rho_{\theta+\epsilon}$ as
\begin{equation}
I_{\mathrm{Fisher}}(\rho_\theta)=4\left(\left.\frac{\partial
d_{\mathrm{Bures}}\left(\rho_{\theta},\rho_{\theta+\epsilon}\right)}{\partial\epsilon}\right|_{\epsilon=0}\right)^{2}\,. 
\end{equation} 
The Bures distance between two quantum
states $\rho_1,\rho_2$ is defined as
\begin{equation}
d_{\mathrm{Bures}}(\rho_{1},\rho_{2})=\sqrt{2}\sqrt{1-\sqrt{F(\rho_{1},\rho_{2})}} \label{eq:Bures}
\end{equation}
where
$F(\rho_1,\rho_2)=(\tr(\sqrt{\rho_1}\rho_2\sqrt{\rho_1})^{1/2})^2$ denotes
the  fidelity between the two states. 
In \cite{Scutaru1998} it was found that for two arbitrary single-mode Gaussian
states $\rho_{1},\rho_{2}$ of the form \eqref{Wg},
\begin{widetext}
\begin{equation}
F(\rho_{1},\rho_{2}) =
\frac{2\exp\left[-\frac{1}{2}\boldsymbol{\Delta}\mathbf{X}^{\top}\left(\boldsymbol{\Sigma}_{1}+\boldsymbol{\Sigma}_{2}\right)^{-1}\boldsymbol{\Delta}\mathbf{X}\right]}{\sqrt{\left|\boldsymbol{\Sigma}_{1}+\boldsymbol{\Sigma}_{2}\right|+\left(1-\left|\boldsymbol{\Sigma}_{1}\right|\right)\left(1-\left|\boldsymbol{\Sigma}_{2}\right|\right)}-\sqrt{\left(1-\left|\boldsymbol{\Sigma}_{1}\right|\right)\left(1-\left|\boldsymbol{\Sigma}_{2}\right|\right)}}
\label{eq:FGauss}
\end{equation}
\end{widetext}
where
$\boldsymbol{\Delta}\mathbf{X}=\langle\mathbf{X}_{1}-\mathbf{X}_{2}\rangle$  
is the mean relative displacement.
Under a smoothness hypothesis, necessary for any CR bound, we have
$\left.\frac{\partial F\left(\rho_{\theta},\rho_{\theta+\epsilon}\right)}{\partial\epsilon}\right|_{\epsilon=0}=0$
and
\begin{equation}
I_{\mathrm{Fisher}}(\rho_\theta)=-2\left.\frac{\partial^{2}F\left(\rho_{\theta},\rho_{\theta+\epsilon}\right)}{\partial\epsilon^{2}}\right|_{\epsilon=0}\,.
\end{equation}

After a straightforward but long and tedious expansion of the fidelity to 
second order we find 
\begin{equation}
I_{\mathrm{Fisher}}(\rho_\theta)=\frac{1}{2}\frac{\tr\left[\left(\boldsymbol{\Sigma}_{\theta}^{-1}\boldsymbol{\Sigma}_{\theta}^{\prime}\right)^{2}\right]}{1+P_{\theta}^{2}}+2\frac{P_{\theta}^{\prime2}}{1-P_{\theta}^{4}}+\boldsymbol{\Delta}\mathbf{X}_{\theta}^{\prime\top}\boldsymbol{\Sigma}_{\theta}^{-1}\boldsymbol{\Delta}\mathbf{X}_{\theta}^{\prime}\,.
\label{eq:IFisherGS} 
\end{equation}
Eq.(\ref{eq:IFisherGS}) shows that the quantum Fisher information depends on
three terms representing the information carried by
\begin{enumerate}
\item the evolution of the noise properties of the state encoded in
  $\bm{\Sigma}_\theta$  
\item the evolution of the purity $P_\theta$ with $\theta$
\item the ``speed'' of displacement $\boldsymbol{\Delta}\mathbf{X}_\theta'=d\langle \mathbf{X}_{\theta+\epsilon}-\mathbf{X}_\theta\rangle/d\epsilon|_{\epsilon=0}$ of
  the state in phase space.  
\end{enumerate}
Equation (\ref{eq:IFisherGS}) provides a generalization of the result for
pure Gaussian single-mode states \cite{Pinel2012} and constitutes the main result of this paper. The second term vanishes
if for the value of $\theta$ under consideration the state is pure,
$P_\theta=1$, under the condition that the eigenvalues of $\rho_\theta$ are 
differentiable at   
that value of $\theta$.

{\em Unification of previous partial results.}
We now show that one obtains from \eqref{eq:IFisherGS} previous partial
results for 
particular measurements. We recall that a general single-mode Gaussian state
can 
always be represented as a squeezed displaced thermal state $\nu$
\cite{Weedbrook2012}, 
\begin{equation}
\rho = R(\psi)D(\alpha)S(\xi)\nu S(\xi)^\dagger D(\alpha)^\dagger R(\psi)^\dagger\,.\label{rhoG}
\end{equation}
where $R(\psi)=\exp(\ii\psi a^\dagger a)$ is the rotation operator, $D(\alpha) = \exp(\alpha a^\dagger - \alpha^* a)$ is the displacement operator and $S(\xi) = \exp(\frac{1}{2}\xi^2 a^{\dagger 2} -\frac{1}{2}\xi^{*2} a^2)$ is the squeezing operator.

The five real parameters can be interpreted physically as:
\begin{itemize}
\item the shift of the state along the $x$ quadrature, parameterized by a $\mathbb R \ni\alpha>0$, and  the phase of the rotation $\psi\in\mathbb R$ 
\item a complex squeezing parameter $\xi = r \e^{\ii\chi}$,
  $r,\chi\in\mathbb R$, where $r>0$ defines the amount of squeezing,  
  and $\chi$ the squeezing direction; we will also use the parameter
  $\sigma=\e^{-r}$ 
\item the purity of the initial thermal state $\nu$, $P_0=1/(2N_{\mathrm{th}}+1)$ 
  where $N_{\mathrm{th}}=\tr(\nu a^\dagger a)$ denotes the number of
  thermal photons. Since squeezing and shifting are unitary operations, the second term in (\ref{eq:IFisherGS}) only contributes if $\theta$
  is a function of $N_{th}$. Otherwise, we have $P_\theta=P_0$.
\end{itemize}

With these parameters, $\boldsymbol{\Delta}\mathbf{X}=2\alpha(\cos\psi,\sin\psi)$, and the
general covariance matrix can then be written as 
\cite{Weedbrook2012} 

\begin{widetext}
\begin{equation}
\boldsymbol{\Sigma} = (2N_{\mathrm{th}}+1) \left( \begin{array}{cc}
\displaystyle{\sigma^2\cos^2(\chi+\psi)+\frac{1}{\sigma^2}\sin^2(\chi+\psi)} & \displaystyle{\frac{1}{2} (\sigma^2-\frac{1}{\sigma^2})
  \sin(2\chi+2\psi) }\\ 
\displaystyle{\frac{1}{2} (\sigma^2-\frac{1}{\sigma^2}) \sin(2\chi+2\psi)}
& \displaystyle{\frac{1}{\sigma^2}\cos^2(\chi+\psi) + \sigma^2\sin^2(\chi+\psi)} 
\end{array} \right)\label{eq:Sigma}
\end{equation}
\end{widetext}

Applying Eq.~(\ref{eq:QCR}), we
find the following expressions for the quantum Fisher information $I_\theta$
for 
all five parameters $\theta\in\{\alpha,\psi,\sigma,\chi,N_{\rm th}\}$ (we
replace from now on the subscript ``Fisher'' with the parameter(s) $\theta$
to be 
varied). 

The quantum Fisher information for the estimation of
$\alpha$ reads
\begin{equation}\label{Ia}
I_\alpha=4 P_0  \left(\frac{1}{\sigma^2}\cos^2(\chi) + \sigma^2\sin^2(\chi)\right)\,.
\end{equation}
Note that amplitude estimation is directly related to the measurement of the
power of the 
electro-magnetic signal. 
As to be expected, $I_\alpha$ is maximal when the state is amplitude
squeezed. For an unsqueezed state, $\sigma=1$, we have 
$I_\alpha=4 P_0$, which for a pure state, $P_0=1$, agrees with the result that one may obtain directly from the overlap of two coherent
states.

The quantum Fisher information for phase estimation reads
\begin{equation}
I_\psi=4 P_0 \alpha^2 \left(\sigma^2\cos^2(\chi) + \frac{1}{\sigma^2}\sin^2(\chi)\right) + \frac{1}{1+P_0^2}\frac{(1-\sigma^4)^2}{\sigma^4}\,.\label{eq:Ipsi}
\end{equation}
The first term depends on the mean field. It is largest when the state is
phase-squeezed i.e.~when $\chi = 0$ and $\sigma >1$. 
The second term depends only on the squeezing-dependent noise properties of the
state and its purity. Each of these two terms corresponds exactly to the
results of \cite{Aspachs2009} where the authors analyze displaced thermal states and
thermal squeezed states. Eq.(\ref{eq:Ipsi}) generalizes these results to the
most general single-mode Gaussian states that can be both squeezed and
displaced at the same time.  
From a metrological perspective the most important
property of $I_\psi$ is its scaling with the mean photon number related to
the displacement, $N=\alpha^2$
\cite{Caves1981,Giovannetti2004,Giovannetti2006}. We see that for large
$\alpha$ 
the first term dominates, and leads for large $N$ to the so-called ``shot noise limited scaling'', $\delta\psi\propto 1/\sqrt{N}$. However the limit can be well below the shot noise limit for large squeezing, and can in principle be arbitrarly small, as strongly squeezed states have large mean energies. For a given total mean photon number $N$ of the state, one can show that the limit on $\delta\psi$ scales as $N^{-3/4}$ \cite{Caves1981,Barnett2003}.

For the estimation of squeezing,  we find asymptotically the same bound as
\cite{Milburn1994,Chiribella2006} for pure Gaussian states. The quantum Fisher
information for the estimation of $\sigma^2$ reads
\begin{equation}
I_{\sigma^2}=\frac{1}{1+P_0^2} \frac{1}{\sigma^4}\,.
\end{equation}
On the other hand, the quantum Fisher information $I_r$ for the squeezing
parameter $r$ is a constant,
which generalizes the result in \cite{Milburn1994}.

The quantum Fisher information relevant for estimating the squeezing angle
is  
\begin{equation} \label{Ichi}
I_\chi=\frac{1}{1+P_0^2} \frac{(1-\sigma^4)^2}{\sigma^4}\,.
\end{equation}
Interestingly, both the
squeezing and its angle can be estimated with a 
sensitivity that reaches, for large $N_{\rm th}$, a constant independent of
$N_{\rm th}$. 
This is in contrast to the
estimation for the thermal 
photon number itself, for which the sensitivity keeps getting worse with larger
photon number. The corresponding quantum Fisher information reads 
\begin{equation}
I_{N_{\rm th}}=\frac{1}{N_{\rm th} + N_{\rm th}^2}\,.
\end{equation}
This can be understood as a consequence of increasing thermal
smearing of the state as function of temperature which leads to larger and
larger (thermal) photon number fluctuations. 
Alternatively, we have the quantum Fisher information for the estimation for
purity 
$I_P=1/(P^2-P^4)$, as follows also from $I_{N_{\rm th}}$ by the laws of
error-propagation.

Eq.(\ref{eq:IFisherGS}) can also be applied to the estimation of other
relevant physical parameters through different parametrizations of the Gaussian state, such as e.g.~the estimation of losses.  Taking as  
initial state 
an amplitude squeezed state with real amplitude $\alpha_0$ and variance
$\sigma^2$ in the amplitude quadrature ($\psi=\chi=0$),  the
amplitude and the covariance matrix of the state read, after an attenuation
of $\eta$, respectively, 
\begin{eqnarray}
\alpha(\eta) &=& \sqrt{1-\eta}\,\alpha_0\,,\\
\boldsymbol{\Sigma}& =& \left( \begin{array}{cc}
\sigma^2 + \eta(1-\sigma^2) & 0 \\
0 & \frac{1}{\sigma^2} + \eta(1-\frac{1}{\sigma^2})
\end{array}\right)\,.
\end{eqnarray}
The quantum Fisher information for the estimation of $\eta$ is found to be
\begin{eqnarray}
I_\eta&=&\frac{1}{1-\eta}\\
&&\times\left(\frac{\alpha_{0}^{2}}{\sigma^{2}+\eta(1-\sigma^{2})}+\frac{(1-2\eta(1-\eta))(1-\sigma^{2})^{2}}{2\eta(2\sigma^{2}+\eta(1-\eta)(1-\sigma^{2})^{2})}\right)\nonumber
\label{eq:Ieta}
\end{eqnarray}
This corresponds exactly to the result of \cite{Monras2007}, if we translate
$\eta$ to the parameter $\phi$ in that paper, as $1-\eta = \cos^2(\phi) =
\e^{-\gamma t}$ where $\gamma$ denotes the rate in the Lindblad master equation
and $t$ the evolution time in the channel.

{\em Extension to multiple parameters.}
In the case of the simultaneous measurement of several parameters
$\bm{\theta}=\theta_1,\ldots,\theta_p$, the quantum Cram\'er--Rao bound
generalizes to a matrix inequality bounding the covariance matrix $\bm{\gamma}$
of the estimators, defined through its matrix elements $\gamma_{ij}=\langle
\theta_i\theta_j\rangle-\langle \theta_i\rangle \langle \theta_j\rangle$. A
lower bound of this matrix is given by the inverse of the quantum Fisher matrix
$\bI(\bthe)$ \cite{Paris2009},
\begin{equation} \label{QCRmulti}
\bm{\gamma}\ge \frac{1}{Q}\bI(\bthe)^{-1}\,.
\end{equation}
The inequality is to be understood in the sense that $\bA\ge \bB$ is equivalent
to $\bA-\bB$ being a positive semi-definite matrix. The quantum Fisher matrix
$\bI(\bthe)$ is defined through the symmetric logarithmic derivative
$L_{\theta_i}$ of the state with respect to a parameter $\theta_i$,
\begin{equation} \label{eq:HFisher}
\bI(\bthe)_{ij}=\tr\left(\rho_{\bm{\theta}}\frac{L_{\theta_i}L_{\theta_j}+L_{\theta_j}L_{\theta_i}}{2}\right)=\tr(\partial_{\theta_i}\rho_{\bm{\theta}}L_{\theta_j})\,.
\end{equation}
The symmetric logarithmic derivative can be expressed in terms of the spectral
decomposition of the density matrix, $\rho_{\bm{\theta}}=\sum_n
\rho_n(\bm{\theta})|\psi_n(\bthe)\rangle\langle\psi_n(\bthe)|$, as
\begin{equation} \label{eq:L}
L_{\theta_i}\equiv 2\sum_{nm}\frac{\langle\psi_m|\partial
  _{\theta_i}\rho_{\bm{\theta}}|\psi_n\rangle}{\rho_n+\rho_m} |\psi_m\rangle\langle\psi_n|\,.
\end{equation}
The sum is over all terms with $\rho_n+\rho_m\ne 0$. Contrary to the
single-parameter case, the bound \eqref{QCRmulti} may not necessarily be 
achievable. In 
the case of a diagonal quantum Fisher information matrix one gets back result
\eqref{eq:QCR}. 
The quantum Fisher matrix defines a
Riemannian metric with metric tensor $g_{ij}$
\cite{Bengtsson2006,Paris2009},
\begin{equation} \label{eq:BuresRiem}
d^2_{\rm Bures}(\rho_{\bm{\theta}},\rho_{\bm{\theta}+d\bm{\theta}})= g_{ij}d\theta_i
d\theta_j=\frac{1}{4}I_{ij}(\bthe)\,. 
\end{equation}
This implies that we can calculate the quantum Fisher matrix by differentiating
the Bures distance $d_{\rm Bures}(\rho_{\bm{\theta}},\rho_{\bm{\theta}+d\bm{\theta}})$ with respect
to the two parameters $d\theta_i$ and $d\theta_j$.

When applying this procedure to \eqref{eq:Bures} with \eqref{eq:FGauss} for the
fidelity of a single-mode Gaussian state, we obtain the matrix element
$I_{\theta_i\theta_j}$ for measuring two parameters $\theta_i$ and $\theta_j$,
\begin{align}I_{\theta_i\theta_j} &
=\frac{1}{2}\frac{1}{1+P_{\bm{\theta}}^{2}}\tr\left(\boldsymbol{\Sigma}_{\bm{\theta}}^{-1}
 \frac{\partial\boldsymbol{\Sigma}_{\bm{\theta}}}{\partial\theta_i}
\boldsymbol{\Sigma}_{\bm{\theta}}^{-1}
 \frac{\partial\boldsymbol{\Sigma}_{\bm{\theta}}}{\partial\theta_j}\right)\nonumber\\   
 & \quad+\frac{2}{1-P_{\bm{\theta}}^{4}}\frac{\partial
 P_{\bm{\theta}}}{\partial\theta_i}
\frac{\partial
 P_{\bm{\theta}}}{\partial\theta_j}\nonumber\\  
 & 
\quad+\left(\frac{\partial\boldsymbol{\Delta}\mathbf{X}}{\partial\theta_i}\right)^{\top}\boldsymbol{\Sigma}_{\bm{\theta}}^{-1}\left(\frac{\partial\boldsymbol{\Delta}\mathbf{X}}{\partial\theta_j}\right)
\,.\label{Iij}
\end{align}
Compared to \eqref{eq:IFisherGS} we see that squared derivatives with respect to
the same parameter $\theta$ are simply replaced by mixed derivatives with
respect to $\theta_i$ and $\theta_j$, such that the diagonal matrix elements
agree with \eqref{eq:IFisherGS}, $I_{\theta_i\theta_i}=I_{\theta_i}$. With
the general expression (\ref{Iij}) one can explicitly calculate the entire
quantum Fisher information matrix with dimension up to $5\times 5$. 

In terms of the parameters introduced in (\ref{rhoG}), there are only two
independent non-vanishing off-diagonal matrix-elements, 
\begin{eqnarray}
I_{\chi\psi}&=& I_{\chi}\\
I_{\alpha\psi}&=& 2P_0 \alpha \left(\frac{1}{\sigma^2} - \sigma^2\right)\sin(2\chi) \,.\label{Iapsi}
\end{eqnarray}

The first equation can be easily understood from (\ref{eq:Sigma}), where
$\chi$ and $\psi$ always appear in linear combination. From \eqref{Iapsi} we see that for
states without displacement ($\alpha=0$), without squeezing ($\sigma=1$), or
squeezing in direction $\chi=0$ the off-diagonal matrix-element
$I_{\alpha\psi}$ vanishes, implying that in this case 
amplitude and phase estimation can be optimized independently and their
individual Cram\'er-Rao bounds reached, no matter how large the initial
thermal photon number is. 
   
Another useful example  of the possibility of statistically independent
measurements is the simultaneous measurement of attenuation $\eta$ and
phase 
$\psi$ when light in an interferometer passes through a phase shifter. A
realistic phase shifter, such as a thin piece of glass, will indeed not only
shift the phase, but typically also lead to some attenuation of the signal
and thus to a mixed state if one does not keep track of the photon number,
such that \eqref{eq:HFisher} applies. 
This situation was considered recently in \cite{Crowley2012}, albeit
for a state with a fixed number of photons, in which case the corresponding
$2\times 2$ quantum Fisher
information matrix is diagonal. Here we see that the same independence holds
for all single-mode Gaussian states. 

{\em In summary}, we have derived the quantum Cram\'er--Rao bound for the
measurement of the five parameters characterizing a general mixed single-mode
Gaussian state of light. Our analysis generalizes and unifies several existing
approaches for particular states or particular single-parameter measurements
\cite{Twamley1996,Paraoanu1998,Scutaru1998,Monras2007,Aspachs2009}.
We have 
also derived the quantum Fisher information matrix that gives a
matrix-valued lower 
bound on the covariance matrix of estimators in the case of the simultaneous
measurement of several parameters, and found that the only two joint
measurements which are generically not independent are those of the phase
together with the amplitude or together with the phase of the squeezing. 
Our results constitute a complete solution
of the problem of the best possible sensitivity for the measurement of an
arbitrary parameter of the most general (not necessarily pure) single-mode
Gaussian state. 

{\em Acknowledgements:} The research is supported by the ANR project Qualitime and the ERC starting grant Frecquam (PJ, NT, CF) and by the Australian Research Council Centre of Excellence for Quantum Computation and Communication Technology, project number CE110001027 (OP). CF is a member of the Institut Universitaire de France.

When completing this manuscript, we became aware of a very recent alternative
approach for estimation of a single parameter with general multimode
Gaussian light by Monras \cite{Monras2013}. While our results agree with the
general 
eq.(13) in that paper when specialized to the single-mode 
single-parameter case, our eq.(\ref{eq:IFisherGS}) contains an extra term
compared to his eq.(16) 
due to the variation of the purity with the parameter. 

\bibliography{biblioQCR}

\end{document}